# Correlating Pedestrian Flows and Search Engine Queries

Vassilis Kostakos, Simo Hosio, Jorge Goncalves
Department of Computer Science and Engineering
University of Oulu
{vassilis, simo.hosio, jgoncalv}@ee.oulu.fi

**ABSTRACT**
An important challenge for ubiquitous computing is the development of techniques that can characterize a location vis-a-vis the richness and diversity of urban settings. In this paper we report our work on correlating urban pedestrian flows with Google search queries. Using longitudinal data we show pedestrian flows at particular locations can be correlated with the frequency of Google search terms that are semantically relevant to those locations. Our approach can identify relevant content, media, and advertisements for particular locations.

**Author Keywords**
Urban flows, online searching, correlation.

**INTRODUCTION**
An important challenge for ubiquitous computing is the development of systems that can sense and characterize their environment. Beyond low-level physical sensing, we are interested in building situated systems that can be aware of their environment and its people. Archetypal examples of such systems include situated displays [7] that "glean" information from the environment and present relevant content. A relevant application domain is pervasive advertising [9], which entails providing people in a particular context appropriate information, services and media.

Here we demonstrate a novel approach for developing situated systems. So far, work on situated systems and pervasive advertising has required human involvement at all stages of development and deployment. For instance, situated displays require explicit or implicit feedback from people in the environment to identify their interests and preferences, which can be subsequently moderated. Similarly, most pervasive advertising techniques rely on traditional marketing approaches to collect information and identify relevant content and services.

The technique we present here relies on the automated collection of longitudinal urban mobility data, specifically data on pedestrian flows. This can be collected using either traditional ticketing mechanisms or using proximity technologies such as Bluetooth and WiFi. Here we demonstrate how such data can be used in conjunction with the service "Google Correlate" (http://www.google.com/trends/correlate) to characterize the environment from which the pedestrian flow data was collected. We show that it is possible to identify semantically relevant Google search queries whose trends correlate with the pedestrian flow data collected for a particular location. The search queries themselves can subsequently be used to identify relevant media such as images and documents, as well as relevant advertisements.

**RELATED WORK**
Google has progressively developed a variety of tools to enable access to aggregated online web search query data. "Google Trends" and "Google Insights for Search" are two real-time systems which provide temporal and spatial activity for a given query. Their main purpose is to model over time the relative popularity of a particular search query. Building on these tools, "Google Flu Trends" provides estimates of Influenza-like Illness activity in the United States, using models based on query data. This tool relies on a model of the relationship between influenza symptoms and a set of web search queries. Using this model it is possible to assess in real time the volume of influenza symptoms based on the web search volume for particular queries.

The tool we use in our work, "Google Correlate" builds on the above tools previously released by Google. The service is effectively a generalization of the "Google Flu Trends" in that it allows for automated query selection across millions of candidate queries for any temporal or spatial pattern of interest. "Google Correlate" is an online system and can provide results in real time. Effectively, this tool can analyze a particular temporal dataset of activity to identify relevant web search queries whose temporal frequency correlates well with the dataset.

What evidence supports that web search queries correlate with real world phenomena? Web search queries have indeed been shown to be useful proxies for estimating many real-world phenomena. For example, they have been shown to correlate with health phenomena including influenza [6], acute diarrhea and chickenpox [11], and salmonella [2]. They also correlate with economic phenomena including movie box office sales, computer game sales, music billboard ranking [5], retail sales including cars, homes, and travel [3], investor attention [4], and claims for unemployment [1]. It is therefore plausible to hypothesize that web searches may also correlate with pedestrian flows and people's patterns of visiting locations across a city.

**STUDY**
Over a period of three years (2009-2012) we collected pedestrian flow data in urban environments using Bluetooth scanning [10]. Our data collection consisted of dedicated Bluetooth scanners that constantly seek discoverable Bluetooth devices in the environment. Previous work has shown that approximately 10% of the population carry discoverable Bluetooth devices [8], but this can vary across demographics and contexts. Our equipment was installed in an enclosure and deployed in various locations across the city. The locations we observed are listed below.

- **University**. This was an indoor location on campus, at a central gathering place nearby the main food court. Due

to the layout of our campus, we expect most people going for lunch or moving across campus to pass by our equipment.

- **Market**. This was an outdoor location in the city centre. It is an open square with benches next to the sea. It is one of the most popular tourist attractions of our City and is busy particularly during summers.
- **Library**. Our equipment was installed in the lobby area of the city library. All visitors of the library pass by our equipment.
- **Sports hall**. Our equipment was installed in the lobby area of a major sports arena which has indoor sports and main events. All visitors of the sports hall pass by our equipment.
- **Shopping street.** This outdoor location is on a shopping street with big nearby shops, about 200 meters away from the market area.

**Correlation with Google search queries**

For each location we calculated the number of unique devices detected for each week of the study, and then used Google Correlate to identify queries from our country (i.e. the country where we collected the pedestrian flow data). Using our pedestrian flows as a target, the analysis provides a long list of search queries whose frequency fluctuates in a similar pattern over time. These results were then ordered in terms of their R-correlation metric. Each query in these results represents an expression that may appear on its own or as part of a longer more complicated query. For instance, the search term "Italian" refers to all search queries including that word, such as "Italian food", "Italian restaurant", and "funny Italian jokes".

Next, we conducted a study with 9 residents. Each participant was given the list of the locations with a description, and the top 10 search queries for each location. Participants rated on a 5-point scale the relevance of the "phrases" to the location itself, and gave additional explanation for their rationale. We did not give any explanation as to how we obtained these "phrases" or what our research is about. We asked the participants to consider the broader context of a particular location, as well as the organizations and institutions that possibly managed the locations. We also instructed them to use online searching in case they were unfamiliar with any of the queries. All residents had first-hand experience of all locations, had visited all of them, and had lived in the city for more than 8 years.

**RESULTS**

Table 1 shows for each location the relevant search queries we identified, the number of *unique* devices detected during our study, the R score for the correlation between the pedestrian flow and search query data, and the average semantic relevance rating reported in our qualitative study. The interrater agreement (Fleiss' kappa) across all results was 0.33 suggesting modest agreement between participants about the relevance of the search queries to the respective locations. The scatterplot in Table 1 shows the relationship between the mean semantic relevance and interrater agreement for each location, suggesting that participants agreed about irrelevant queries but disagreed about relevant queries.

Two further correlations are shown in Figure 1, which shows the "research" query for the University and "bikes" query for the Shopping street locations respectively. The pedestrian flow (blue) and search query (red) data is normalized and then correlated.

**DISCUSSION**

There are many reasons why, in principle, our methodology should not produce meaningful results. But it does. The qualitative ratings and our own analysis suggest that our methodology can identify search queries that are semantically relevant to the locations, along with some irrelevant queries. Many results do not appear to be random but are indeed semantically relevant to the locations. The relevance of the results varies across locations in our study, with certain locations giving rise to more semantically relevant queries than other locations, as shown in the scatterplot in Table 1.

There are two main types of criticism that colleagues have expressed when reviewing our findings. The first line of criticism has to do with the observation that our correlations are not causal relationships, but rather likely to be affected by external confounding variables such as the weather and seasonal variations. If indeed seasonal variations affect both pedestrian flows and the search queries, then can our findings be of value? A second type of criticism has to do with the fact that while the pedestrian flows were collected from geographically confined locations, the search query data is aggregated at a national level due to technical restrictions of Google Correlate. How is it possible, then, that the correlations we obtain can be meaningful? We address both types of criticism next.

**Can the results be valid if they are affected by confounding variables?**

Our analysis indicates that, despite criticism, many search queries identified in our results are semantically relevant to the locations we studied. We do believe that weather and seasonal variations play an important role in the correlation process, but this does not appear to hinder our analysis. The most substantial evidence supporting this claim is a comparison of the data form the Market and the Shopping street. These are two outdoor locations approximately 200 meters apart from each other. If indeed only seasonal variations were a dominant factor, then we would expect that our data from these two locations to be similar. After all, the weather at these locations is near-identical.

However, our methodology produced substantially different search queries for the Market and Shopping street. Our analysis for the Market location produced search queries referring to outdoor plazas, typical summer phenomena, and other popular tourist attractions. On the other hand, the Shopping street location, although also outdoor, produced search queries generally related to cycling, e.g. bike shops, bike repair, etc. This location hosts the busiest bike rack area downtown (and there are indeed bike shops nearby). Similarly, our three indoor locations produced quite distinct flows and queries, even though they are all indoor locations and we would expect homogeneous seasonal variations for indoor locations.

Therefore, we argue that while seasonal variations exist in our data, they do not seem to have a significant impact on our results. Rather, we believe that weather and seasonal effects are "situationally appropriated". In other words, the fact that the weather is hotter during summer means that more people google for busy outdoor locations or cycling products during the summer, and more people actually visit such locations during summer. As such, our methodology is able to identify for each particular location the way in which people change their behavior *in situ* as seasons vary.

**How can location-bound data correlate with national-level data?**

It is challenging to reliably interpret why location-bound pedestrian flows should correlate with search query data collected at a national level. We iterate the fact that our results *are* in fact meaningful and we *are* able to identify semantically relevant queries for different locations. There are two reason why we believe the mismatch in granularity between our pedestrian flows and search query data is not hampering our analysis.

First, while our pedestrian flow data is location-bound, it represents a non-negligible part of the population of our country. For instance, our data from the Shopping street includes almost 120,000 unique devices, potentially representing 2.2% of the county's population (although we admit that not all unique devices necessarily represent a unique person). Thus we believe that because all locations we studied are population attractors with tens of thousands of unique visitors, the fine temporal variations in our data generate a unique footprint which becomes salient in the correlation process. Therefore, even though millions of potential queries are compared against our pedestrian flow data, only relevant ones result in high correlation scores.

However, this explanation does not account for the fact that our findings suggest that when more people are searching for "bikes" on a national level, more people visit the particular Shopping street in our city. This relationship makes us hypothesize that our results actually demonstrate a relationship between national search queries and "location motifs". The particular locations in our study just happen to be instances of these "location motifs". We argue that when more people are searching for "bikes" across the country, more people visit shopping areas across the country, one of which happens to be the Shopping street in our study. This hypothesis also has a corollary: if we conduct a similar study across multiple Libraries or Markets or Universities, we expect to find strong correlation patterns between these locations of similar "motif", and we expect to identify similar search queries for these locations respectively.

| | R | Relevance (1-5) | | R | Relevance (1-5) | | R | Relevance (1-5) |
|---|---|---|---|---|---|---|---|---|
| **University campus** (10891 devices) | | | **Sports arena** (23603 devices) | | | **Library** (17720 devices) | | |
| markkinointi (marketing) | 0.8596 | 3.00 | petri matikainen (an ice hockey coach) | 0.8134 | 2.67 | otimaa24 (a news blog) | 0.7764 | 2.11 |
| ergonomia (ergonomics) | 0.8430 | 1.56 | tervahölkkä (a local sports event) | 0.7937 | 3.22 | puheenvuoro (a news blog) | 0.7722 | 2.56 |
| SFS (a standards organisation) | 0.8399 | 2.11 | terwamaraton (a local sports event) | 0.7709 | 3.56 | mail | 0.7642 | 3.00 |
| organisaatio (organization) | 0.8361 | 3.44 | stafettkarnevalen (sports event) | 0.7584 | 2.33 | tryg (an insurance company) | 0.7639 | 1.22 |
| tutkimus (research) | 0.8345 | 4.56 | judgement day | 0.7575 | 1.11 | garderob | 0.7634 | 1.33 |
| asetus (setting) | 0.8339 | 1.89 | isojano (a beer festival) | 0.7548 | 2.11 | iphone 3 | 0.7625 | 1.33 |
| management | 0.8334 | 2.56 | lejon (lion / ice hockey team symbol) | 0.7534 | 1.67 | puukello (wooden clock) | 0.7606 | 1.33 |
| projekti (project) | 0.8315 | 3.78 | glasvegas (a band) | 0.7503 | 1.00 | kaikki mitäraksin (a novel) | 0.7560 | 2.78 |
| rakenne (structure) | 0.8308 | 2.67 | koivu (an ice hockey player / birch) | 0.7476 | 2.56 | xvideo | 0.7556 | 1.56 |
| industry | 0.8300 | 2.78 | end of world | 0.7463 | 1.00 | outlook 2010 | 0.7552 | 1.89 |
| **Market** (83641 devices) | | | **Shopping street** (118846 devices) | | | | | |
| hietsun kirppis (flea market) | 0.9239 | 2.33 | bikes | 0.6456 | 4.00 | | | |
| punkin (mite's) | 0.9110 | 1.11 | kompressori (compressor) | 0.6398 | 1.00 | | | |
| punkki (mite) | 0.9087 | 1.11 | ormax | 0.6392 | 1.00 | | | |
| samppalinna (a tourist attraction) | 0.9087 | 1.67 | peuramaa golf (a golf center) | 0.6300 | 1.11 | | | |
| suomenlinna (a tourist attraction) | 0.9081 | 1.78 | pyöräkorjaamo (bicycle repair shop) | 0.6298 | 2.11 | | | |
| purema (bite) | 0.9080 | 1.11 | bike shop | 0.6249 | 1.78 | | | |
| hietaniemen kirpputori (a flea market) | 0.9080 | 1.89 | kansalaisen (citizen) | 0.6247 | 2.44 | | | |
| punkin purema (mite bite) | | | ringside | 0.6242 | 1.22 | | | |
| hietalahden kirpputori (a flea market) | 0.9079 | 1.22 | shimano (bike parts manufacturer) | 0.6240 | 3.00 | | | |
| tammisaari (a city) | 0.9070 | 2.33 | jani kaasalainen (an online bike store) | 0.6193 | 1.22 | | | |
| | 0.9015 | 2.00 | | | | | | |

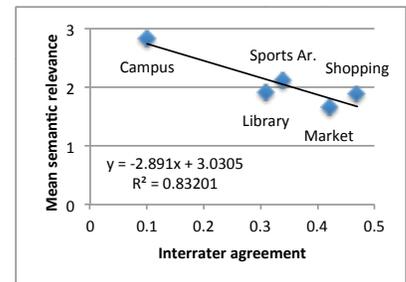

Table 1. For each location: total unique devices detected, top 10 search queries based on the R-correlation metric, average semantic similarity assigned by human raters. The scatterplot shows for all locations the relationship between average semantic similarity for top 10 queries and interrater agreement (Fleiss' kappa).

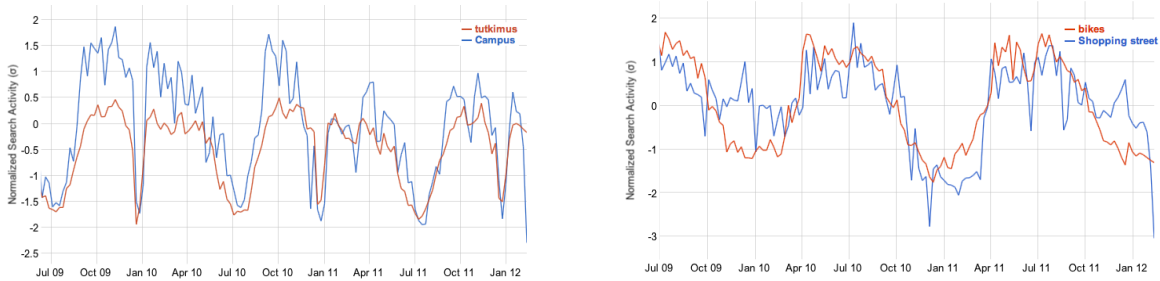

Figure 1. Correlation between pedestrian flows (red) and search query volume (blue).
Left: data from the University Campus. Right: data from the Shopping street.

We actually tested this hypothesis by obtaining WiFi traces from two distinct high schools in our city. The data was analyzed using the same method as described here. The search queries across both locations were semantically similar, and include search terms that relate to our country's educational system and educational tools. This finding is encouraging, and warrants further investigation.

**Implications for situated systems**

We describe a methodology to "convert" pedestrian flow data into phrases that are semantically relevant to particular areas and contexts. This approach can have a substantial impact on how we design and implement contextual and situated ubiquitous systems. These keywords can be used to identify relevant media, services, data, and advertisements. One straightforward mechanism to implement this is to use web search tools to identify relevant media. For instance, using Google's "image search" with the queries our technique identifies, we are able to retrieve location images that are semantically relevant to the locations in our study. Similarly, our approach could be used for pervasive advertising purposes. For example Google's advertising tools can be used to retrieve advertisements relevant to the keywords identified in our analysis.

Finally, we note that an understanding of the locations in question is required to asses the semantic relevance of the search queries produced by our method, and to distinguish them from noise. For this reason, our method requires human moderation. Our participant feedback also shows that when humans are not fully briefed about these results and their context, they find it hard to identify semantic relationships. Particularly, the scatterplot in Table 1 shows that participants agreed about *irrelevant* queries, but disagreed about what is *relevant*. Our own heuristic analysis suggests that in many cases a semantic relationship becomes salient when the context of the results is known.

**CONCLUSION AND FUTURE WORK**

This paper is one of the first to present empirical evidence that urban pedestrian flows correlate with semantically relevant search engine queries. Our work opens a promising avenue for building situated systems and pervasive advertising techniques that can identify relevant media, information and services by sensing the environment around them.

The work we have presented here is not the result of an engineered effort, but rather of a phenomenon that we have come across in our analysis of urban data. In many ways we are still quite far from fully understanding the underlying mechanisms that are manifesting as the correlations we have identified. Our so far limited understanding is in no way a show-stopper. We argue that the method we have presented can already be used as a mechanism for building contextual and situated systems and services.